\documentclass[10pt,letterpaper]{article}
\usepackage{opex3}
\usepackage{cite}

\begin{document}

\title{Design of an optical reference cavity with low thermal noise and flexible thermal expansion properties}

\author{J. Zhang, Y. X. Luo, B. Ouyang, K. Deng, Z. H. Lu$^\ast$, and J. Luo}

\address{Key Laboratory of Fundamental Physical Quantities Measurement of Ministry of Education \\School of Physics, Huazhong University of Science and Technology \\1037 Luoyu Road, Wuhan, P. R. China 430074}

\email{zehuanglu@mail.hust.edu.cn} 



\begin{abstract}
An ultrastable optical reference cavity with re-entrant fused silica mirrors and a ULE spacer structure is designed through finite element analysis. The designed cavity has a low thermal noise limit of $1\times10^{-16}$ and a flexible zero crossing temperature of the effective coefficient of thermal expansion (CTE). The CTE zero crossing temperature difference between a composite cavity and a pure ULE cavity can be tuned from $-10\ ^{\circ}$C to $23\ ^{\circ}$C, which enables operation of the designed reference cavity near room temperature without worrying about the CTE zero crossing temperature of the ULE spacer. The design can be applied to cavities with different lengths. Vibration immunity of the cavity is also achieved through structure optimization. 
\end{abstract}



\section{Introduction}
Ultra-stable lasers \cite{prl1999_Young} are essential in high-resolution laser spectroscopy \cite{prl2005_Oskay}, optical frequency standards \cite{science2001_Hg, science2004_Sr, prl2005_Yb, prl2006_Sr, prl2007_Al, oc2007_Wang, science2008_Sr, nature2008_Sr, prl2009_Yb}, gravitational wave detection \cite{cqg2008_Willke}, and fundamental tests of physics \cite{science2008_Al, prl2008_Sr}. To achieve such a frequency stabilized local oscillator, the laser is typically servo locked to a high-finesse reference cavity by using Pound-Drever-Hall (PDH) technique \cite{apb1983_PDH}. Within the servo locking bandwidth, the stability of the locked laser is determined by the stability of the optical path-length of the reference cavity. At low frequencies (below 100 Hz) the most severe fluctuations of the cavity length are caused by environmental disturbances such as seismic and acoustic noise. At higher frequencies these disturbances are usually damped by active vibration isolation. It can be shown that the sensitivity of the reference cavity to the low frequency environmental disturbances can be minimized by optimizing cavity geometry and mounting method \cite{ol2005_Notcutt, pra2006_Chen, apb2006_Nazarova, pra2007_Webster, pra2008_Hansch, pra2009_Millo, oe2009_Zhao, oc2010_Zhao}. 

While the sensitivity of a reference cavity to mechanical vibrations can be suppressed such that the influence of these perturbations can be ignored, there still exists a fundamental thermal noise limit to the level of stability achievable with the reference cavity \cite{prl2004_Numata}. This limit is attributed to Brownian motion of the reference cavity spacer, mirror substrates, and optical coatings \cite{prl2004_Numata}. As have been widely reported before \cite{oc2002_Nevsky, pra2006_Ye, ol2007_Ye, pra2008_Gill, pra2009_Knoop}, using short length cavities ($\sim10$ cm) constructed entirely from ULE glass, the relative frequency stability characterized by Allan deviation was limited by thermal noise to the level of $10^{-15}$. Several methods can be used to lower the thermal noise limit, such as increasing the reference cavity spacer length, increasing the laser spot size, and using high mechanical $Q$ materials. Of these methods, using high $Q$ materials offer the biggest improvements. 

To lower the thermal noise limit of reference cavities, a commonly used high $Q$ material is fused silica (FS). Because FS has a much larger coefficient of thermal expansion (CTE) ($500$ ppb/K) than that of ULE ($\pm 50$ ppb/K) \cite{oe2009_Fox}, it is not appropriate to use FS as cavity spacer material. Since the dominant thermal noise contribution comes from the mirror substrates, it would be more suitable to just use FS as mirror substrates, and keep ULE as cavity spacer material. Replacing the ULE mirrors of a short ULE cavity by FS mirrors can lower the thermal noise limit approximately by a factor of two \cite{pra2008_Gill}. Combining with a longer length cavity spacer, it is then possible to reach a thermal noise limit of $10^{-16}$.

However, the CTE mismatch between FS and ULE can cause a problem. Temperature changes inside the reference cavity can create stress which bends the mirror substrates, and introduces a larger effective CTE for the composite cavity. This problem can be alleviated by setting the operating temperature of the reference cavity at the zero crossing temperature of the effective CTE. If a ULE spacer has its zero crossing temperature of CTE at room temperature, then the zero crossing temperature of the effective CTE will be much lower. However, this is not very convenient, because cooling a reference cavity is technically more challenging than simply heating the reference cavity. If the temperature is set too low, it can introduce extra problems, like water condensation on cavity housing, \textit{etc.} For practical purpose, it is preferable to stabilize the reference cavity temperature a little bit above room temperature. 

Presently there are several approaches to achieve this objective. One is to find a ULE piece that has negative CTE at room temperature, and in combination with positive CTE of FS, it is possible to make the zero crossing temperature of the effective CTE to be above room temperature \cite{oe2009_Fox}. Another method is to contact ULE rings to the back side of FS mirrors to prevent the mirror bending due to CTE mismatch \cite{josab2010_Sterr}. As a consequence, the zero crossing temperature of the effective CTE can be above room temperature. Both methods are used by Jiang \textit{et al.}, and a fractional frequency instability of $2\times10^{-16}$ was achieved \cite{nature2011_Jiang}.

The above two approaches have their limitations. The first approach relies on the availability of ULE material with negative CTE at room temperature. This requires the good will of the ULE manufacturer, and might not be always available. The second approach in principle can be used to tune the zero crossing temperature over a range of about 30 $^{\circ}$C, but it only works for a smaller diameter ULE spacer \cite{josab2010_Sterr}. To lower the thermal noise of the reference cavity, it is common to choose a longer cavity length with bigger diameter. The tuning range then is limited to only about $5\sim6$ $^{\circ}$C for this geometry according to our calculations. Normal ULE has the zero crossing temperature of CTE varying between 5 $^\circ$C and 35 $^\circ$C \cite{oe2009_Fox,Corning_ULE}, so with this limited tuning range it is not always possible to raise the zero crossing temperature of the effective CTE to above room temperature.

Another approach is to use a re-entrant mirror design for the reference cavity \cite{eftf2010_Gill}. By using a re-entrant mirror design, the direction of FS mirror thermal expansion is opposite to that of ULE spacer, so in principle they can be compensated better, and introduce less stress to the mirrors. In reality the tuning range of the zero crossing temperature difference is about 20 $^\circ$C $\sim$ 30 $^\circ$C. Consequently, this cavity design is only suitable for cases when ULE spacers have a zero crossing temperature lower than 5 $^\circ$C. In this paper, we analyze a modified re-entrant mirror design for the reference cavity, where the ULE rings for the re-entrant FS mirrors are replaced with FS rings. With this design modification, the zero crossing temperature of the effective CTE can be tuned to much greater range, and it is very easy to design a composite cavity so that the zero crossing temperature of the effective CTE is above room temperature. 

Since a low thermal noise reference cavity design is central to this paper, a detailed discussion of the thermal noise limit of a ULE cavity is presented in Sec. $2$. Based on our calculation results, a particular geometry of the reference cavity is chosen, and re-entrant FS mirrors are used to replace the traditional ULE mirrors. With this geometry, in Sec. $3$ we analyze the zero crossing temperature of the composite cavity with finite elements analysis (FEA), and compare the results with that of the other approaches. For completeness, in Sec. $4$ we also design the cavity to be vibration-insensitive. Finally we concluded in Sec. $5$.

\section{Calculation of Thermal Noise Limits}
The stability of a reference cavity is limited by an inevitable mechanical thermal fluctuation due to the Brownian motions of materials. From the fluctuation-dissipation theorem, the most significant parameters in the calculation of the thermal noise spectrum for a given structure are the mechanical quality $Q$ factors of materials. Materials with higher $Q$ factors generate less mechanical dissipation, thus contribute smaller thermal noise. To get an order of magnitude estimation of their contributions, using the formulas given in \cite{prl2004_Numata}, the power spectral density of mirror displacement due to thermal noise, $G_m(f)$, which typically dominates the spacer contribution, is given by
\begin{equation}
G_m(f)=\frac{4k_BT}{2\pi f}\frac{1-\sigma_{sub}^2}{\sqrt{\pi}E_{sub} w_0}\frac{1}{Q_{sub}}(1+\frac{2}{\sqrt{\pi}}\frac{1-2\sigma_{sub}}{1-\sigma_{sub}}\frac{Q_{sub}}{Q_{coating}}\frac{d}{w_0}),
\label{Gm}
\end{equation}
where $k_B$ is the Boltzmann's constant, $T$ is the temperature, $\sigma_{sub}$ is the Poisson's ratio of mirror substrate, $E_{sub}$ is the Young's modulus of mirror substrate, $w_0$ is the beam radius, $Q_{sub}(Q_{coating})$ is the mechanical quality factor of the mirror substrate (coating), and $d$ is the coating thickness ($\sim5$ $\mu$m). In addition, the spacer contribution can be written as
\begin{equation}
G_{sp}(f)=\frac{4k_BT}{2\pi f}\frac{L}{3\pi R^2E_{sp}}\frac{1}{Q_{sp}},
\label{Gsp}
\end{equation}
where $L$ and $R$ are the spacer length and radius, respectively, $E_{sp}$ and $Q_{sp}$ is the Young's modulus and mechanical quality factor of the spacer, respectively. The total reference cavity displacement power spectral density due to thermal noise is
\begin{equation}
G_t(f)=2G_m(f)+G_{sp}(f).
\end{equation}
From which we can calculate the Allan deviation of the relative stability of the cavity as
\begin{equation}
\sigma_y=\frac{\sqrt{2\ln2f}\cdot\sqrt{G_t(f)}}{L}.
\label{allan}
\end{equation}

From Eq.~(\ref{Gm}) and Eq.~(\ref{Gsp}), we can see that by choosing materials of higher $Q$ factors, the thermal noise limit can be efficiently lowered. Among the low CTE materials that are commonly used in ultra-stable cavities, ULE has a $Q$ factor of $6.1\times10^4$ that is more than one order of magnitude larger than that of Zerodur ($3.1\times10^3$). Furthermore, FS has an even larger $Q$ factor of about $10^6$ \cite{Numata}. In cases where thermal noise limits are the ultimate limiting factor of achievable cavity stability, it is very effective to use a higher $Q$ factor material (like FS) as mirror substrate while keeping ULE as spacer material.

\begin{figure}[htb]
\centering\includegraphics[width=8cm]{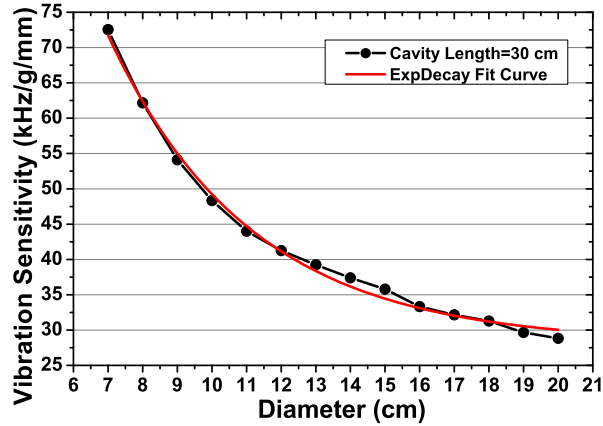}
\caption{Variation of vibration sensitivity of a 30 cm long cavity with respect to spacer diameter. The vibration sensitivity becomes much smoother at diameter larger than 15 cm. Consequently, a spacer diameter of 15 cm is chosen in all following simulations}
\label{Vibration}
\end{figure} 

According to Eq.~(\ref{allan}), using longer cavity can also have the benefit of suppressing the thermal noise. In addition, the total thermal noise limit also depends on the laser beam sizes on the reflecting mirror surfaces. Larger beam radius contributes a lower thermal noise. Taking into account of all these effects, we calculated the thermal noise limits of cavities with different dimensions and configurations, expressed in both Allan deviation and frequency spectral density, and the data are listed in Table~\ref{Table}. By replacing ULE mirrors with FS mirrors and increasing the cavity length from 10 cm to 30 cm, the thermal noise limit can be suppressed by nearly one order of magnitude. Consequently in the following we will consider the design of a long, FS mirror substrate cavity. Notice here that for longer cavity, the spacer radius is also increased. This is due to the fact that longer cavity is more susceptible to vibration influence, and larger cavity can help to suppress this effect. The result is shown in Fig.~\ref{Vibration}. The detail is explained in Sec. $4$.

\begin{table}[htbp]
\caption{Thermal noise limits of optical reference cavities with different configurations and dimensions. $L$, $R$, and $w_{01}/w_{02}$ are the spacer length, spacer radius and approximate mode spot sizes (intensity $1/e^2$ radius) on the plano-confocal cavity mirrors, respectively. $\sigma$ is the Allan deviation of cavity stability and $\sqrt{S_{\nu}(1Hz)}$ is the root cavity frequency noise spectral density at 1 Hz for the 1064 nm light.}
\label{Table}
\renewcommand\arraystretch{1.5}
\begin{center}
\begin{tabular}{cccccc}\hline

Mirror Substrate &$L$   &$R$    &$w_{01}/w_{02}$ &$\sigma$    &$\sqrt{S_{\nu}(1Hz)}$ \\ 
                 &(cm)  &(cm)   &($\mu$m)        &            &(Hz/$\sqrt{Hz}$) \\\hline
ULE              &10    &3.5    &260/291         & 7.1$\times10^{-16}$    &  0.170 \\
FS               &10    &3.5    &260/291         & 3.8$\times10^{-16}$    &  0.090 \\
ULE              &30    &7.5    &394/471         & 1.8$\times10^{-16}$    &  0.043 \\
FS               &30    &7.5    &394/471         & 8.4$\times10^{-17}$    &  0.020 \\\hline             

\end{tabular}
\end{center}
\end{table}

\section{Compensation of Thermal Distortion}

By replacing the ULE mirrors with FS mirrors, CTE mismatch between the two materials will cause structure deformation when environmental temperature changes, and thus change the effective CTE of the composite cavity, $\alpha_{eff}$, and zero crossing temperature of the effective CTE, $T_0(eff)$. To tune this zero crossing temperature to near room temperature, strategies include attaching ULE rings to FS mirrors, and using re-entrant FS mirrors inside the spacer \cite{josab2010_Sterr, eftf2010_Gill}. In this section, both strategies are analyzed with FEA for a 30 cm long reference cavity with a diameter of 15 cm. Based on the analysis, we make modifications to the design and also give the simulation results. 

In our calculations, we assume the ULE spacer has length of $L$, and optical contacted FS mirrors have radius of $R$. Due to the CTE difference, a temperature variation $dT$ results in a difference of the radial expansion between $dR=(\alpha_{FS}-\alpha_{ULE})RdT$. This results in an axial expansion $dA$ of the mirrors if we assume the optical contact to be perfectly rigid. Assuming a linear
stress-strain relation, the radial expansion $dR$ and the axial expansion $dA$ are related by a constant $\delta$ to be $dA=\delta dR$. The effective cavity CTE is then
\begin{eqnarray}
\alpha_{eff}&=&\frac{\alpha_{ULE}LdT+2dA}{LdT}\nonumber\\
            &=&\alpha_{ULE}+2\delta\frac{R}{L}(\alpha_{FS}-\alpha_{ULE}).
\label{effective}
\end{eqnarray}
We are primarily interested in the difference of zero crossing temperatures, 
\begin{equation}
\Delta T=T_0(eff)-T_0(ULE),
\end{equation}
between the effective CTE of cavity and the original CTE of ULE material. We aim to find a structure design which can tune the effective zero crossing temperature of the cavity CTE, $T_0(eff)$, to be slightly higher than room temperature (25 $^\circ$C $\sim$ 28 $^\circ$C) for better thermal control of the system. Since normal ULE material has zero crossing temperature of CTE, $T_0(ULE)$, in the range between 5 $^\circ$C and 35 $^\circ$C \cite{oe2009_Fox,Corning_ULE}, this means that the tuning ability of an optimal design should at least achieve $-10\ ^\circ\mathrm{C}\leq \Delta T\leq 23\ ^\circ\mathrm{C}$.

\begin{figure}[htb]
\centering\includegraphics[width=12cm]{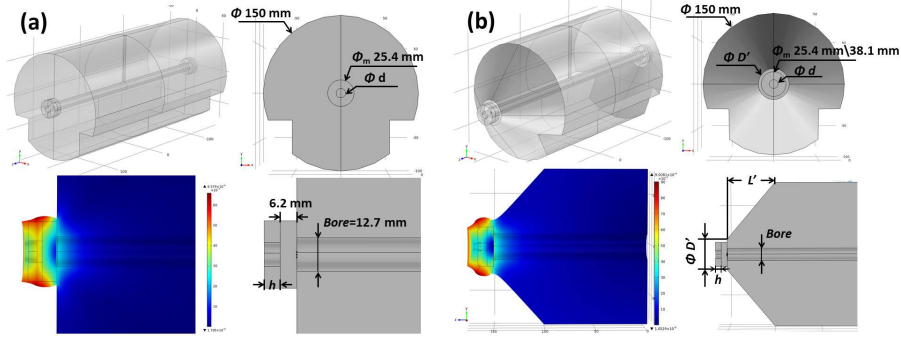}
\caption{Geometries of sandwich cavities and FEA simulations of the deformations caused by thermal expansion mismatch when temperature is increased by 1 $^\circ$C. (a) cavity with a cylindrical spacer. (b) cavity with a tapered spacer. Color scale shows the axial displacement. $d$ and $h$ are the inner diameter and thickness of the ULE rings, respectively. The notch structure is designed for vibration suppression purpose, and will be explained in Sec. $4$.}
\label{Zhangf1}
\end{figure}

\begin{figure}[htb]
\centering\includegraphics[width=12cm]{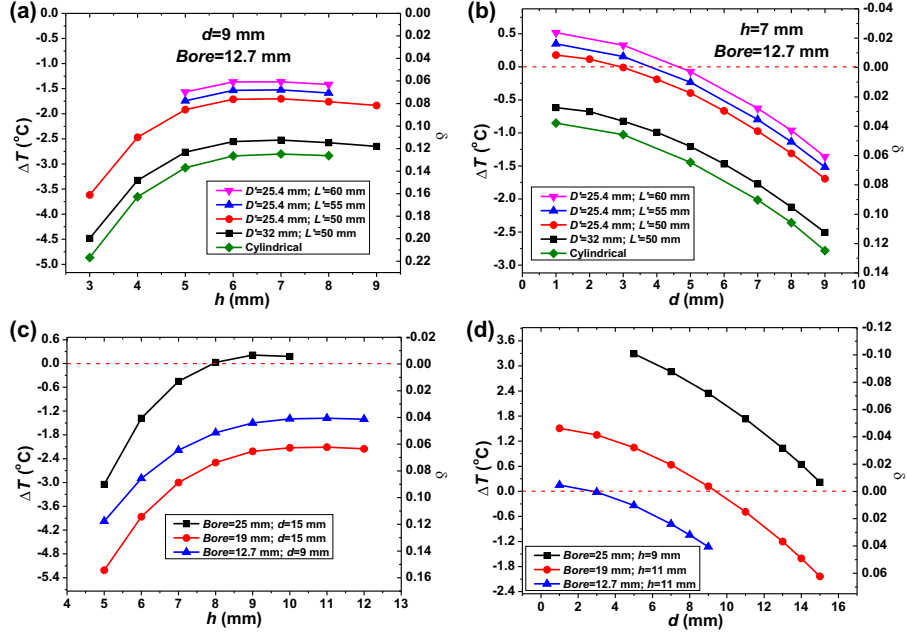}
\caption{FEA simulations of the CTE zero crossing temperature tuning capabilities of sandwich cavities with different dimensions and configurations, including cylindrical and tapered structures (upper panel) as well as designs of 1.5 inch mirror structures (lower panel). $d$ and $h$ are the inner diameter and thickness of the ULE rings, respectively, other parameters of $D'$ and $L'$ are defined in Fig.1. Right vertical axis shows the coupling coefficient $\delta$ as defined in Ref.~\cite{josab2010_Sterr} that directly connects the axial mirror displacement with the radial expansion between mirror and spacer.}
\label{Zhangf2}
\end{figure}

Following Ref.~\cite{josab2010_Sterr}, we add ULE rings to the back side of FS mirrors to form a ``sandwich'' structure and analyze the thermal properties of the combined cavity using both Comsol multiphysics and ANSYS FEA packages. Both analyses give consistent results. Numerical analysis indicates that the temperature difference $\Delta T$ to be only slightly higher than zero for a 30 cm long cavity, even if many efforts have been paid to optimize the cavity structures, including the introduction of a tapered structure to reduce the areas of the ends of spacer and the adoption of larger mirrors (1.5 inch diameter). Cavity geometries and FEA analysis results are shown in Fig.~\ref{Zhangf1} and Fig.~\ref{Zhangf2}, respectively. As shown in Fig.~\ref{Zhangf2}, for cavities with simple cylindrical structures, it is impossible to tune the temperature difference $\Delta T$ to be greater than zero. For tapered cavities, an upper limit of 3 $^\circ$C of temperature difference in the case of 1.5 inch mirrors structure can be achieved. Although this might be good enough for some cases when the ULE spacer has a $T_0(ULE)$ near or higher than room temperature, it is not flexible enough because you are not normally able to choose the very batch of material with the characteristics you want. It should be noticed that although smaller inner diameter of the ULE ring $d$ can further increase the range of $\Delta T$, it has a practical limit and should not be too small, otherwise the entrance of the laser beam will be affected, and the scattered light caused by laser diffraction at small hole will introduce unwanted noise.

\begin{figure}[htb]
\centering\includegraphics[width=12cm]{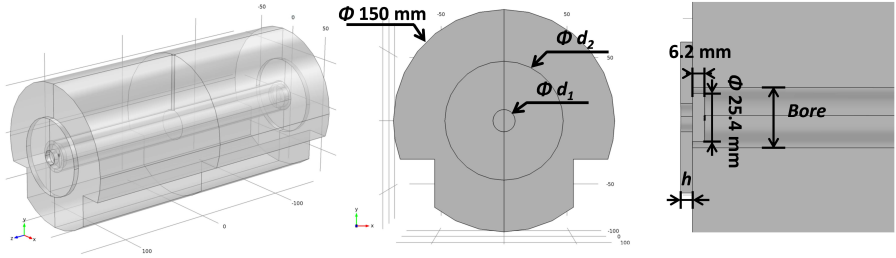}
\caption{Geometry and dimensions of cavity with re-entrant FS mirrors optically contacted to ULE rings. The ULE ring has an inner diameter of $d_1$, an outer diameter of $d_2$, and $h$ is the thickness of rings.}
\label{Zhangf3}
\end{figure}

\begin{figure}[htb]
\centering\includegraphics[width=12cm]{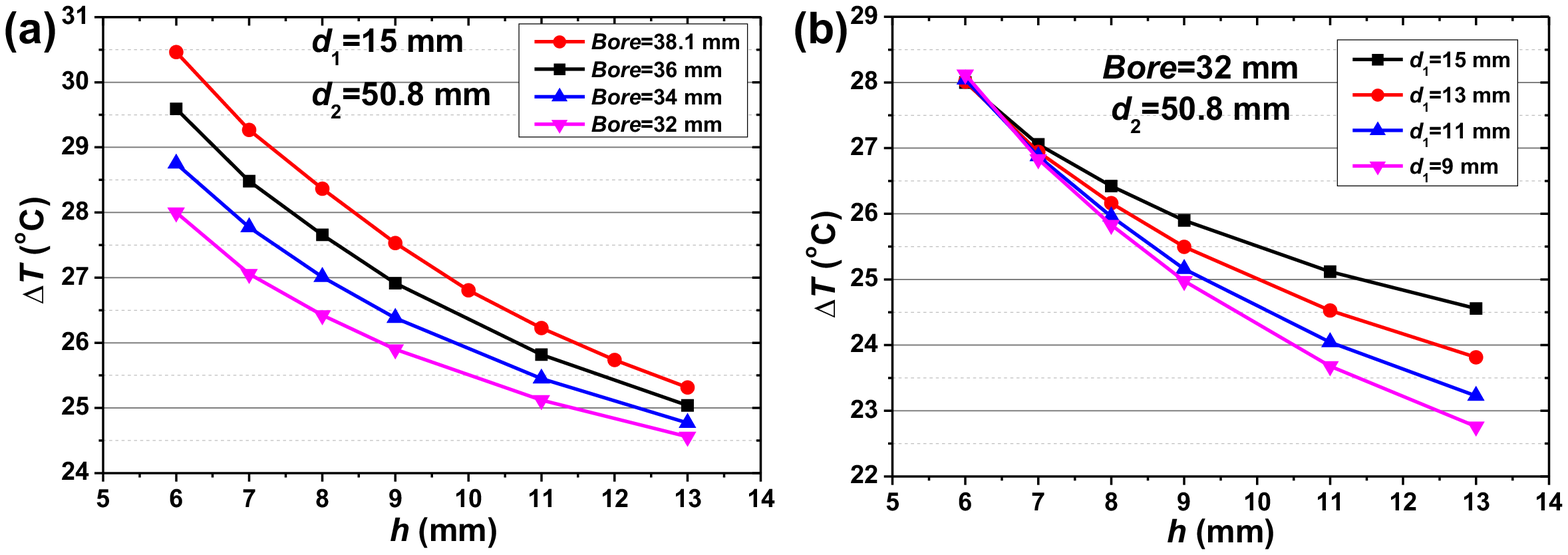}
\caption{FEA simulations of the CTE zero crossing temperature tuning range of re-entrant cavities with FS mirrors optically contacted to outer ULE rings with different structure parameters. $d_1$ and $d_2$ are the inner and outer diameters of ULE rings, respectively, and $h$ is the thickness of rings. Definitions of these dimensional parameters are shown in Fig.~\ref{Zhangf3}.}
\label{Zhangf4}
\end{figure}

Cavities with FS mirrors re-entrant within the bores of ULE spacers have the advantage of compensation of opposite thermal expansion between spacers and mirrors inside. Geometry of the cavity is illustrated in Fig.~\ref{Zhangf3}. In this structure, FS mirrors are optically contacted to ULE rings which are in turn optically contacted to the ends of ULE spacer. Simulations show that temperature difference between $T_0(eff)$ and $T_0(ULE)$ can be as large as 30 $^\circ$C, and the results are shown in Fig.~\ref{Zhangf4}. Again, the inner diameter of the ring $d_1$ cannot be too small to affect the entrance of laser beam. In contrast with that of the ``sandwich'' structure, it is difficult to tune the $T_0(eff)$ lower. The tuning range of the zero crossing temperature difference, $\Delta T$, is about 20 $^\circ$C $\sim$ 30 $^\circ$C. Consequently, this cavity design is only suitable for cases when ULE materials have a $T_0(ULE)$ lower than 5 $^\circ$C. 

\begin{figure}[htb]
\centering\includegraphics[width=12cm]{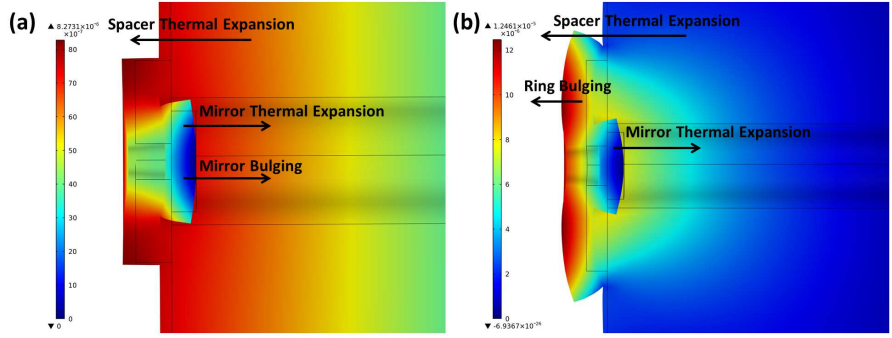}
\caption{FEA simulations of the deformations caused by CTE mismatch of re-entrant cavities; (a) deformation between FS mirror and ULE ring causing the mirror bulges in the same direction as its thermal expansion direction; (b) deformation between FS ring and ULE spacer causing the ring bulges in the opposite direction of the mirror thermal expansion direction. Color scale shows the axial displacement.}
\label{Zhangf5}
\end{figure}

One of the reasons for such a high temperature difference is due to huge difference of CTE between FS and ULE. FS has a CTE of about $5.2\times10^{-7}$ K$^{-1}$ at room temperature while ULE has CTE of less than $10^{-8}$ K$^{-1}$ (depending on temperature $T$) \cite{oe2009_Fox,Corning_ULE}. However, the more important reason is the deformation of mirrors caused by CTE mismatch. This deformation makes the FS mirrors bulge in the same direction of its thermal expansion direction, as shown in Fig.~\ref{Zhangf5}(a), leading to a result that only long cavities are suitable for this design. This process can be understood theoretically using Eq.~\ref{effective}. For the re-entrant cavity structure, the effective CTE of the cavity can be expressed as:
\begin{eqnarray}
\alpha_{eff}&=&\frac{\alpha_{ULE}L-2\alpha_{FS}h_m}{L-2h_m}-2\delta\frac{R}{L}(\alpha_{FS}-\alpha_{ULE})\nonumber\\
            &\approx&\alpha_{ULE}-2\alpha_{FS}\frac{h_m}{L}-2\delta\frac{R}{L}(\alpha_{FS}-\alpha_{ULE}),
\label{effCTE1}
\end{eqnarray}
where $h_m$ is the thickness of the FS mirror substrates, and is assumed to be $6.2$ mm in our calculations. Here we have used the assumption that $L >> h_m$. In first order approximation we can write the CTE of ULE around its zero crossing temperature $T_0(ULE)$ to be
\begin{equation}
\alpha_{ULE}(T)=\beta(T-T_0(ULE)),
\label{beta}
\end{equation}                                                                                                    
where $\beta$ is a linear temperature coefficient and is taken to be $2\times10^{-9}$ K$^{-2}$ \cite{josab2010_Sterr}. Thus the effective CTE can be expressed as:
\begin{eqnarray}
\alpha_{eff}&\approx&\beta[T-T_0(ULE)-\frac{2\alpha_{FS}h_m+2\delta R(\alpha_{FS}-\overline{\alpha_{ULE}})}{\beta L}]\nonumber\\
            &\equiv&\beta[T-T_0(eff)].
\label{effCTE2}
\end{eqnarray} 
Here we have used the fact that $2\delta R << L$ to replace the $\alpha_{ULE}$ in the numerator with temperature independent value of $\overline{\alpha_{ULE}}$. The effective zero crossing temperature of CTE of the cavity can then be expressed as
\begin{equation}
T_0(eff)=T_0(ULE)+\frac{2\alpha_{FS}h_m+2\delta R(\alpha_{FS}-\overline{\alpha_{ULE}})}{\beta L}.
\label{deltaT}
\end{equation}                                                      

According to the above equations, to make sure the temperature difference of $\Delta T=T_0(eff)-T_0(ULE)$ to be smaller than 23 $^\circ$C, the spacer length $L$ must be at least 14 cm even without considering the effect of deformation coefficient $\delta$. If $\delta$ is positive, then the mirrors bulge in the same direction of its thermal expansion direction, as shown in Fig.~\ref{Zhangf5}(a), the required cavity length will be even longer, typically longer than 30 cm. Currently spacer length longer than 30 cm is very difficult to obtain, and has higher vibration sensitivity.

\begin{figure}[htb]
\centering\includegraphics[width=8cm]{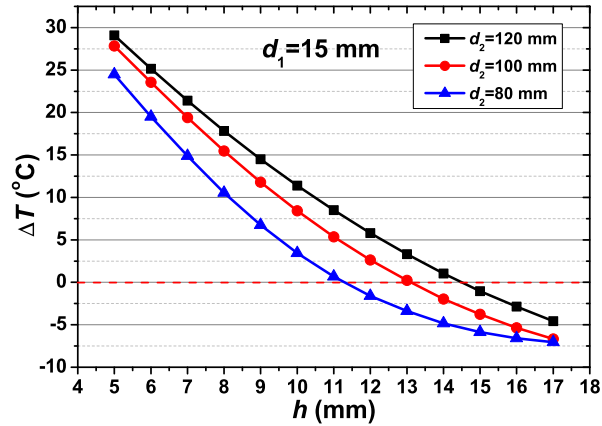}
\caption{Tuning range of CTE zero crossing temperature of re-entrant cavities with FS mirror optically contacted to outer FS rings instead of ULE rings. The definitions of dimensional parameters are the same as those in Fig.~\ref{Zhangf3}.}
\label{Zhangf6}
\end{figure}

To obtain a more flexible tuning range of $\Delta T$ with a moderate cavity length, the coefficient $\delta$ in Eq.~(\ref{deltaT}) needs to be negative to compensate the huge difference between the CTE of FS and ULE. Therefore, a modified design of re-entrant cavity is applied, in which the FS mirrors are optically contacted to FS rings instead of ULE rings. In this configuration, the deformation of CTE mismatch is between the FS rings and ULE spacer, causing the FS rings bulge in the opposite direction of the mirror thermal expansion direction, as shown in Fig.~\ref{Zhangf5}(b). In another word, this cavity structure has a negative $\delta$. Simulations of cavities with this new structure are shown in Fig.~\ref{Zhangf6}.

According to the simulations, the modified re-entrant cavity yields a larger tuning range of $\Delta T$ from -7 $^\circ$C to 25 $^\circ$C simply by varying the thickness $h$ of FS rings, and this range can be adjusted further by altering the outer diameter $d_2$ according to the characteristics of ULE batch used and the set-point temperature of the thermal control system. With this design, the flexibility requirement of $-10\ ^\circ\mathrm{C}\leq \Delta T\leq 23\ ^\circ\mathrm{C}$ is easily achieved. Through comparison, cavities with FS rings of relatively smaller outer diameter $d_2$ are more suitable because of the less area of optical contact as well as the compatibility with vibration immunity design which will be discussed in Sec. $4$. As a result, in the following simulations and analysis, an outer diameter of 80 mm of FS ring is adopted. 

\begin{figure}[htb]
\centering\includegraphics[width=12cm]{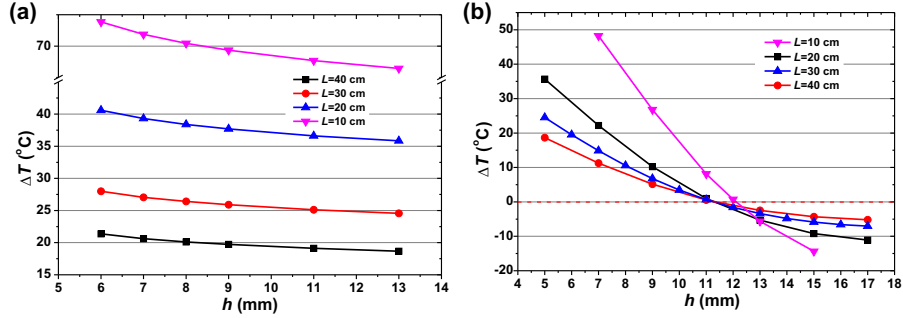}
\caption{Comparison of the zero crossing temperature tuning capabilities for different cavity lengths between the original re-entrant cavity (a) and the modified cavity with FS rings (b). The inner diameters of rings $d_1$ are 15 mm for both cases.}
\label{Zhangf7}
\end{figure}

Figure \ref{Zhangf7} shows the zero crossing temperature tuning capabilities with different cavity lengths between the original and the modified re-entrant cavity. The zero crossing temperature tuning ranges are much narrower for the original design (only about 5 $^\circ$C) than those of the new design (more than 30 $^\circ$C). Furthermore, to obtain a moderate temperature tuning (less than 25 $^\circ$C), the cavity length must be longer than 30 cm with the original design. In comparison, for the new design, variation of ULE spacer lengths from 10 cm to any longer length can achieve a moderate temperature tuning range. It is interesting to note that all curves except $L=10$ cm in Fig.~\ref{Zhangf7}(b) intersect and cross zero at the same position where the ring thickness $h$ is 11.2 mm. Referring to Eq.~(\ref{deltaT}), this intersection with zero reveals the fact that when the ring thickness is at this particular value ($h$=11.2 mm) the thermal expansions of FS mirrors are exactly compensated by the deformation strain of FS rings no matter how long the cavity is. In this case, the effective CTE of the cavity is the same as that of the ULE spacer material. For $L=10$ cm case, the requirement $2\delta R << L$ is not valid, which causes the zero crossing point to be different with that of longer cavities.

\begin{figure}[htb]
\centering\includegraphics[width=12cm]{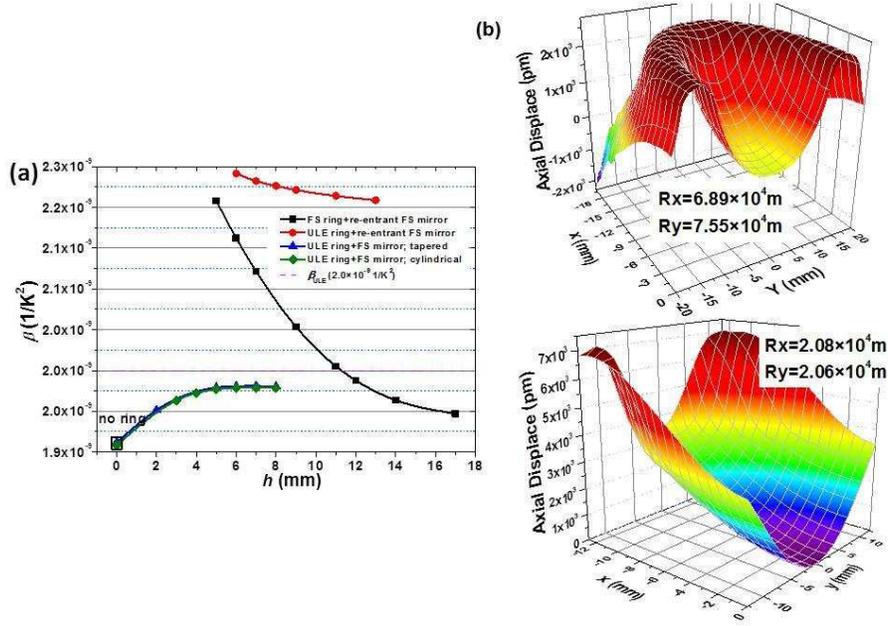}
\caption{Comparison of properties between different designs of composite cavities: (a) linear temperature coefficient $\beta$ of the effective CTE; (b) mirror deformations due to CTE mismatch of sandwich design (upper) and re-entrant designs (below), respectively.}
\label{Zhangf8}
\end{figure}

In this work, the linear temperature coefficient of the effective CTE and mirror deformations of the modified cavity are also studied. The results are shown in Fig.~\ref{Zhangf8} and listed in Table~\ref{Table2}. According to Eq.~(\ref{beta}), the linear temperature coefficient $\beta$ of the effective CTE also depends on cavity structure. Figure \ref{Zhangf8}(a) shows the variation of $\beta$ with respect to ring thickness $h$ of different cavity designs, including the sandwich cavity and the two re-entrant cavities. The coefficients are all near $2\times10^{-9}$ K$^{-2}$, which is the original linear temperature coefficient of ULE material, and the exact values depend on the detailed structure parameters. 

In addition, the deformations of mirror caused by CTE mismatch are also analyzed and are compared between different cavity designs. The results are shown in Fig.~\ref{Zhangf8}(b). Mirrors of the ``sandwich'' design and re-entrant designs all deform in a convex shape at the mirror centers, and the added radius of curvature (negative in these cases) has a magnitude of approximately $10^4$ m to $10^5$ m, which are 4 to 5 orders of magnitude larger than those of the cavity mirrors (0.5 m or 1 m), and thus has negligible contribution. As can be seen from the figure, ``sandwich'' structure causes more abrupt distortions within the mirror. Details of the properties of cavities with different structure designs are listed in Table~\ref{Table2}, including their thermal noise limits. It should be noted here that in the ``sandwich'' design and re-entrant designs, the introduction of ULE or FS rings increases the thermal noise limit by a very small amount. This effect is taken into account in the thermal noise limits calculations in Table~\ref{Table2}.

\begin{table}[htbp]
\caption{Properties of reference cavities with different structure designs. Relative deformation is defined as the ratio between the original mirror radius of curvature and the extra radius of curvature caused by mirror deformations (both horizontal and vertical directions have similar results).}
\label{Table2}
\renewcommand\arraystretch{1.5}
\begin{center}
\begin{tabular}{cccc}\hline

Structure                      &Thermal noise         &$\Delta T$                         &Relative deformation\\ \hline
FS Mirror only                 &$8.4\times10^{-17}$   &$\sim -10$ $^\circ$C               &$5\times10^{-5}$\\
FS Mirror+ULE ring (sandwich)  &$8.5\times10^{-17}$   &$-3\sim 3$ $^\circ$C               &$1\times10^{-5}$\\
FS Mirror+ULE ring (re-entrant)&$8.4\times10^{-17}$   &$20\sim 30$ $^\circ$C              &$5\times10^{-5}$\\
FS Mirror+FS ring (re-entrant) &$8.4\times10^{-17}$   &$-7\sim 25$ $^\circ$C              &$5\times10^{-5}$\\\hline             

\end{tabular}
\end{center}
\end{table}

\section{Design of Vibration Insensitive Cavity} 
The cavities designed with re-entrant FS mirrors and FS rings have an excellent CTE zero crossing temperature tuning property with a very low thermal noise limit. In order to reach this thermal noise limit, the cavity should be further designed to be vibration insensitive. A quantitative analysis of the elastic deformation of reference cavities provides valuable guidance for cavity design. Since the time-dependent vibration perturbation consists of a broad spectrum, in principle a full dynamic analysis of the cavity deformation is required to correctly predict the cavity vibration sensitivity. This dynamic analysis to our knowledge is not available in the literature. However, we can greatly simplify the process by performing static analysis. 

A static analysis can be used to approximate dynamic analysis because the following two conditions are satisfied. First, only the vibration modes with frequency lower than 100 Hz are of interest, because the modes with higher frequencies will be significantly damped by either passive or active vibration isolations used in all reference cavities systems. Second, the considered cavity dimension is much smaller than the cavity vibration characteristic wavelength, meaning that all particles move in-phase inside each eigenmode. Thus the dynamic response of the cavity can be mimicked by applying a static force with the random acceleration frozen at that moment. For example, consider ULE spacer material (with $\rho$ of 2.21 g/cm$^3$ and $E$ of 67.6 GPa), its dispersion relation is \cite{pra2006_Chen}
\begin{equation}
f=\sqrt{\frac{E}{\rho}}\frac{1}{\lambda}.
\end{equation}
For oscillation frequency of 100 Hz, the characteristic wavelength $\lambda$ of 55 m is much greater than the simulated 0.3 m cavity length. Consequently, at these low frequencies considered here the static analysis is a reliable substitute for the dynamic analysis.

In static analysis the elastic deformation of cavities caused by a gravity-like acceleration are studied. To understand the principle, for simplicity we assume the cavity is held in a plane instead of several mounting points. If the cavity is horizontally held on a plane from bottom, compression caused by acceleration normal to the supporting plane will cause expansion in length because of the non-zero Poisson's ratio; on the other hand, if the cavity is hung on the plane from top, the deformation is reversed, the cavity elongates in height and shortens in length. When the cavity is held on its symmetry plane, the lengthening of the upper part and the shortening of the bottom part compensate each other and the axial length of the cavity remains unchanged under vertical accelerations. In practice, cavities are held on discrete points instead of the complete plane, which will cause additional bending deformation as well. The bending effect along with the simple compression or expansion, contribute to the total cavity length variation. In this case, the optimal supporting plane is no-longer the symmetry plane and the cavity length variation under acceleration is dependent on both the discrete supporting positions and the mounting plane of those discrete points.

The bending of cavity will cause mirror tilting as well. If the cavity is supported on points locating at the two opposite sides along the horizontal axial length, it will bend downwards under vertical acceleration and the mirrors at both sides tilt correspondingly; if the cavity is supported on the middle of its length, as an extreme, it will bend upwards, and the mirrors tilt in an opposite way. There must be a critical position along the length of the cavity (Airy points), on which the cavity is supported, so that the mirror tilting is eliminated by the compensation of both bending processes.

\begin{figure}[htb]
\centering\includegraphics[width=10cm]{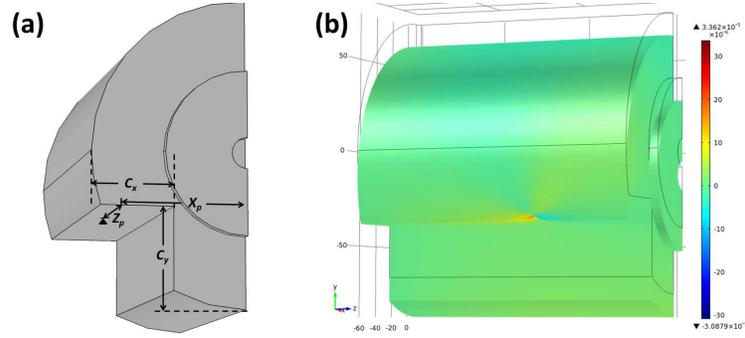}
\caption{(a) Dimensions ($C_x$ and $C_y$) of the notched structure, and positions of mounting supports ($X_p$ and $Z_p$) on the notch surface. (b) FEA simulations of the deformation of a quarter section of the cavity with optimal notch sizes and supporting positions under vertical vibration.}
\label{Vibration1}
\end{figure}

To eliminate both cavity length variation and mirrors tilting simultaneously in the case of our cavity design, we apply a notched structure \cite{pra2006_Chen, apb2006_Nazarova, pra2007_Webster, oe2009_Zhao} to the cavities as shown in Fig.~\ref{Vibration1}. In simulations, we implement FEA analysis by applying a gravity-like force on the cavity vertically with $g=9.8$ m$/\mathrm{s}^2$, and the support with $1\times1$ mm$^2$ square on the notch surface is constrained to simulate the mounting condition. Optimal dimensions of the notched structure are searched so that the two kinds of critical mounting positions with zero length variation and zero mirrors tilting of the cavity overlap, while maintaining a less rigorous requirement of mounting accuracy and machining tolerance. The definitions of parameters about the dimensional and mounting configurations are also illustrated in Fig.~\ref{Vibration1}.

\begin{figure}[htb]
\centering\includegraphics[width=12cm]{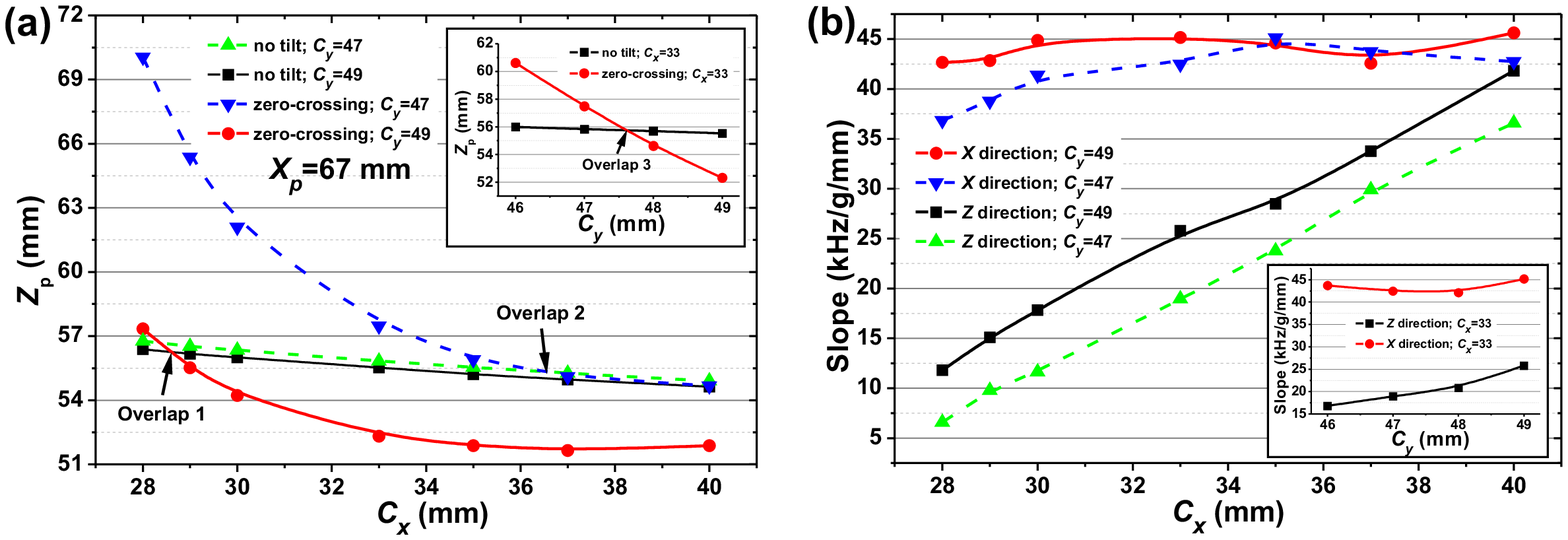}
\caption{(a) Comparison of critical positions, $Z_p$, for cavity mounting with different notch sizes. (b) the slopes near the zero crossing positions of curves related to the dependence of resonance frequency deviation on mounting positions under vertical vibrations with different notch sizes.}
\label{Vibration2}
\end{figure}

\begin{figure}[htb]
\centering\includegraphics[width=12cm]{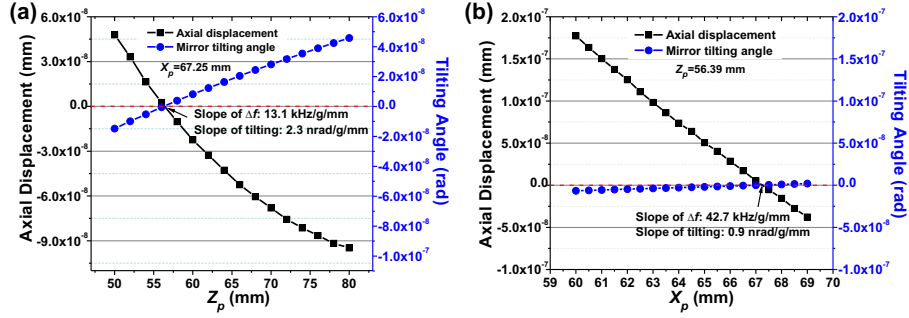}
\caption{(a) FEA simulations of the axial length variations at the mirror center and mirror tilting angles at different mounting positions of $Z_p$. (b) FEA simulations of the axial length variations at the mirror center and mirror tilting angles at different mounting positions of $X_p$. The cavity has notch sizes of $C_x$=28 mm and $C_y$=49 mm, under a gravitational acceleration of $g=9.8$ m$/\mathrm{s}^2$.}
\label{Vibration3}
\end{figure}

During simulations, cavities with different notch sizes ($C_x$ and $C_y$) are examined to search for an optimal dimension with which an overlap of the above two critical mounting positions is available. In most cases, the mounting positions satisfying those two conditions do not overlap, which means only one of the effects can be eliminated. Analysis results are shown in Fig.~\ref{Vibration2} in which the mounting positions with zero length variations and zero mirrors tilting of different notch sizes are compared. Calculations reveal that the mounting positions with zero mirror tilting angles are not sensitive to the changes of the notch sizes, while the zero crossing mounting positions (where the length variations of cavities are immune to vertical vibrations) are much more sensitive to the notch sizes, as shown in Fig.~\ref{Vibration2}(a) and its inset. It is also found that at a particular value of $C_y$ (or $C_x$) there is a corresponding $C_x$ (or $C_y$) that can result in the overlap of these two critical positions. Another important parameter is the slope of the cavity length variation curve, which reflects the sensitivity of vibration immunity to mounting positions accuracy and machining tolerance. Different slopes of cavities with different notch sizes are shown in Fig.~\ref{Vibration2}(b). It indicates that the sensitivity of the frequency variation as mounting positions changes in the transverse ($X$) direction is not sensitive to the notch sizes, neither $C_x$ nor $C_y$, and has a value of approximately 42 kHz/g/mm. In contrast, the sensitivity with mounting positions changes in the axial ($Z$) direction steadily decreases with decreasing notch sizes of both $C_x$ and $C_y$. As a result, smaller notches are much more preferred when optimizing the cavity dimensions. We choose dimensions of $C_x$=28 mm and $C_y$=49 mm and the corresponding optimized vibration sensitivity of the cavity are shown in Fig.~\ref{Vibration3}.

With the optimized notch dimensions, both the length variation and the mirror tilting have a value of zero at the same mounting position. As can be seen in Fig.~\ref{Vibration3}, at the zero crossing point, the frequency variation has a very small slope against mounting position deviation in $Z$ direction, which is 13.1 kHz/g/mm, corresponding to a relative frequency instability of $4.3\times10^{-11}$ /g/mm at a laser wavelength of 1064 nm. The slope of mirror tilting in the axial direction is about $2.3$ nrad/g/mm, corresponding to about $4.5$ kHz/g/mm, thus the relative frequency instability of $1.5\times10^{-11}$ /g/mm if the optical axis is displaced from the mechanical axis by $\pm1$ mm. Comparably, the slope of frequency variation in transverse ($X$) direction is four times larger than that in $Z$ direction, as shown in Fig.~\ref{Vibration3}(b), indicating that the cavity length variation is more sensitive and critical to the accuracy of mounting positions in $X$ direction. The slope of $42.7$ kHz/g/mm corresponds to a relative frequency instability of $1.4\times10^{-10}$/g/mm at $1064$ nm. The sensitivity of mirror tilting to mounting accuracy in $X$ direction is very small ($6.1\times10^{-12}$ /g/mm at $\pm1$ mm optical axis displacement) and can be neglected.

\section{Conclusion}
In conclusion, we have proposed a new ultra-stable optical reference cavity design with ULE spacer, re-entrant FS mirrors and FS rings based on extensive FEA simulations for different cavity structures. The designed cavity has an ultra-low thermal noise limit at the levels of $1\times10^{-16}$. The zero crossing temperature of the effective CTE of the designed cavity can be easily tuned to above room temperature with any ULE spacer materials, as long as the CTE zero crossing temperature of  the selected ULE spacer is measured beforehand. Currently the CTE zero crossing temperature of the ULE spacer can be measured with an accuracy of $\pm 1$ $^\circ$C at reasonable cost (Stable Laser Systems). This accuracy level is more than enough for our design purpose. No matter what the zero crossing temperature of the selected ULE spacer is measured to be, one can always design a cavity with compatible FS rings thickness such that the zero crossing temperature of the effective CTE is above room temperature. The design is applicable for cavities with different lengths. In addition, we have designed the reference cavity to be vibration insensitive through FEA structure optimization.

Ultra-low thermal noise limit reference cavities play a key role in realizing ultra-narrow linewidth lasers. Reference cavities based on ULE and FS materials are still very popular choices for lab-based or portable ultra-narrow linewidth lasers. To further reduce the thermal noise limit to below $1\times10^{-16}$, ULE based reference cavities are no longer adequate. To this end, there are currently strong interests in the developments of cryogenic cavities based on single crystal silicon or sapphire \cite{arXiv_PTB, arXiv_Humboldt}. We hope the technique discussed in this paper can also stimulate ideas in the design of these cryogenic cavities.

\section*{Acknowledgments}
We thank Yan-Yi Jiang and Long-sheng Ma for helpful discussions, and M. Notcutt for technical suggestions. The project is partially supported by the National Basic Research Program of China (Grant No.2012CB821300), the National Natural Science Foundation of China (Grant Number 61108025 and 11174095), and Program for New Century Excellent Talents by the Ministry of Education.

\end{document}